\documentstyle[12pt]{article}
\begin{document}

\begin{center}
{\Large \bf Can the fluctuations of a black hole
be treated thermodynamically?}\\
\vskip .5cm
K. Ropotenko\\
\centerline{\it State Department of communications and
informatization} \centerline{\it Ministry of transport and
communications of Ukraine} \centerline{\it 22, Khreschatyk, 01001,
Kyiv, Ukraine}
\bigskip
\verb"ro@stc.gov.ua"

\end{center}
\bigskip\bigskip

\begin{abstract}
Since the temperature of a typical Schwarzschild black hole is very
low, some doubts are raised about whether the fluctuations of the
black hole can be treated thermodynamically. It is shown that this
is not the case: the thermodynamic fluctuations of a black hole are
considerably larger than the corresponding quantum fluctuations.
Moreover the ratio of the mean square thermodynamic fluctuation to
the corresponding quantum fluctuation can be interpreted as a number
of the effective constituents of a black hole.
\end{abstract}

\bigskip\bigskip

Black holes being exact vacuum solutions of Einstein equations of
the gravitational field are, undoubtedly, geometric objects. On the
other hand, due to quantum effects they acquire thermodynamic
properties. Understanding the statistical origin of black hole
thermodynamics is still a central problem in black hole physics. The
thermodynamic quantities which describe a system in equilibrium are,
almost always, very nearly equal to their mean values. But
deviations from the mean values sometimes occur. They are called
fluctuations. The existence of the fluctuations has important
meaning because it demonstrates statistical nature of the
thermodynamic laws. In the fluctuation theory a distinction is
usually made between, strictly speaking, thermodynamic fluctuations
and quantum ones. The conditions for the fluctuations to be
thermodynamic are \cite{land}

\begin{equation}
\label{cond1} T\gg \hbar/\tau,~~~\tau\gg \hbar/T,
\end{equation}
where $T$ is the temperature of a system and $\tau$ is the
characteristic time of change of a thermodynamic quantity.
Otherwise, when $T$ is too low  or when $\tau$ is too small (a
thermodynamic quantity fluctuates too rapidly) the fluctuations
cannot be treated thermodynamically, and the purely quantum
fluctuations dominate.

Typically the mass of a black hole is considerably larger than the
Planck mass, $M\gg m_P$, and its temperature $T\sim m_P^{2}/M$ is
very low. In addition, for a thermodynamic system $\tau$ is bounded
by the time it takes for a perturbation to propagate through the
system. That is, $\tau \geq R$, where $R$ is the characteristic size
of the system. For a black hole the characteristic size coincides
with its gravitational radius $R_g=2GM$ and $T\sim 1/R_g$. Thus for
a black hole we expect \cite{hod}

\begin{equation}
\label{cond2}T \sim {\hbar/\pi \tau }
\end{equation}
and the violation of the conditions (\ref{cond1}). In this
connection, an important question arises: What is a relation between
thermodynamic and quantum fluctuations of a black hole? In this note
I want to compare quantum fluctuations of a Schwarzschild black hole
with their thermodynamic counterparts.

We begin with thermodynamic fluctuations. As mentioned above, the
black hole is a thermodynamical system with the temperature and
entropy given by
\begin{equation}
\label{temp}T=\frac{1}{8\pi G M}=\frac{1}{4\pi R_g},
\end{equation}

\begin{equation}
\label{ent}S=\frac{\hspace{0cm}\mbox{\emph{Horizon
area}}}{4G}=\frac{\pi R^{2}_{g}}{l^{2}_P}\,,
\end{equation}
which are connected by the first law of black hole thermodynamics in
the form $dE=TdS$, where the internal energy $E$ is identified with
the black hole mass $M$. In thermodynamics the mean square
fluctuations of the fundamental thermodynamic quantities are related
to the specific heat $C$. For a black hole
\begin{equation}
\label{heat}C=T\left(\frac{\partial{S}}
{\partial{T}}\right)=-\frac{1}{8\pi\, G\,T^{2}}=-2S.
\end{equation}
As is well known, a negative heat capacity means that a system
cannot be in stable equilibrium with an infinite reservoir of
radiation at temperature $T$. A well known way to stabilize a black
hole is to place it in a appropriate cavity so that an environmental
heat bath is finite \cite{fro}. In this paper we will assume that
black hole has reached equilibrium with its own radiation, the whole
system being enclosed in the appropriate cavity and we can use a
canonical ensemble. Thus for the mean square fluctuations of the
black hole internal energy (mass), inverse temperature, and entropy
we have respectively:

\begin{equation}
\label{mass1}\langle(\Delta{M})^{2}\rangle_{th}=\frac{\partial(-M)}{\partial\beta}
=\frac{C}{\beta^{2}}\sim m^{2}_P,
\end{equation}

\begin{equation}
\label{temp1}\langle(\Delta{\beta})^{2}\rangle_{th}=-\frac{\partial(\beta)}{\partial
M} =\frac{\beta^{2}}{C}\sim m^{-2}_P,
\end{equation}

\begin{equation}
\label{ent1}\langle(\Delta S)^{2}\rangle_{th}=C=2S,
\end{equation}
where I have added the subscript 'th' to refer to the thermodynamic
fluctuations and omitted signs connected with $C$. As is easily
seen, (\ref{ent1}) can be also rewritten in terms of the
fluctuations in the particle numbers
\begin{equation}
\label{num}\langle(\Delta N)^{2}\rangle_{th} \sim N,
\end{equation}
where
\begin{equation}
\label{num1}N=\frac{\hspace{0cm}\mbox{\emph{Horizon area}}}{l_P^{2}}
\end{equation}
can be interpreted as the number of independent constituents of a
black hole. As is well known, such fluctuations in the particle
numbers inheres in classical particles \cite{land}. Since the mean
square fluctuations (\ref{mass1}) - (\ref{ent1}) don't depend on any
extensive black hole parameter, the meaning of the relative
fluctuations is obviously empty.

Let us now consider quantum fluctuations. To make precise
calculations we need a theory of quantum gravity. But it is still
absent. Despite this we can make some order-of-magnitude estimates.
Suppose, following Frolov and Novikov \cite{fro}, that a fluctuation
in the geometry occurs in a spacetime domain with a characteristic
size $R$ so that the value of the metric $g$ deviates from the
expectation value $\langle g \rangle$ by $\Delta g$. Then, since the
curvature in the domain changes by a quantity of order $\Delta g
/(R^{2}\langle g \rangle)^{2}$, it follows that the change in the
action $I$ of the gravitational field is
\begin{equation}
\label{act} \Delta I \sim \frac{\Delta g \\ R^{2}}{\langle g \rangle
\\ G}.
\end{equation}

The probability of such a quantum fluctuation is considerable if
only $\Delta I \sim \hbar$ so that (\cite{whe}, \cite{har})

\begin{equation}
\label{metr} \frac{\Delta g }{\langle g \rangle }\sim
\frac{l^{2}_P}{R^{2}}.
\end{equation}

This relation being applied to a Schwarzschild black hole determines
the fluctuation of the gravitational radius $R_g$  (\cite{fro},
\cite{yor}, \cite{ford}),

\begin{equation}
\label{flu} \Delta R_g \sim \frac{l^{2}_P}{R_g}.
\end{equation}

From (\ref{flu}) we immediately obtain the mean square fluctuations
of the black hole mass
\begin{equation}
\label{mass2} \langle (\Delta M)^{2}\rangle_{q} \sim
\frac{m^{4}_P}{M^{2}},
\end{equation}
inverse temperature $\beta=1/T$,
\begin{equation}
\label{temp2} \langle (\Delta \beta)^{2}\rangle_{q} \sim
\frac{1}{m^{4}_P \beta ^{2}},
\end{equation}
and entropy
\begin{equation}
\label{ent2} \langle (\Delta S)^{2}\rangle_{q} \sim 1,
\end{equation}
where I have added the subscript 'q' to refer to the quantum
fluctuations. As in the thermodynamic case, we can rewrite
(\ref{ent2}) in terms of the fluctuations in the particle numbers,
\begin{equation}
\label{num2}\langle(\Delta N)^{2}\rangle_{q} \sim 1.
\end{equation}
As is easily seen from (\ref{mass2})-(\ref{ent2}), the relative
quantum fluctuations, as opposed to the thermodynamic case, can have
a meaning, and we can write
\begin{equation}
\label{mass3}\frac{\sqrt{\langle(\Delta{M})^{2}\rangle_{q}}}{M}\sim
\frac{1}{N},
\end{equation}

\begin{equation}
\label{temp3}\frac{\sqrt{\langle(\Delta{\beta})^{2}\rangle_{q}}}{\beta}\sim
\frac{1}{N},
\end{equation}

\begin{equation}
\label{ent3}\frac{\sqrt{\langle(\Delta{S})^{2}\rangle_{q}}}{S}\sim
\frac{1}{N}.
\end{equation}
Consequently, for large $N$ (which is true for a typical black hole)
they are quite negligible.

Of course, quantum fluctuations of macroscopic observables should
obey the uncertainty principle. Let the uncertainty in the value of
the gravitational radius be equal (\ref{flu}); then the
corresponding uncertainty in the value of the momentum is
$\Delta{p}\sim M$. Taking into account (\ref{num}), (\ref{num1}), we
can represent it in the form
\begin{equation}
\label{mom} \Delta{p}\sim \Delta E \sim \Delta N m_P \sim N^{1/2}
m_P\sim M.
\end{equation}
This allows us to regard a black hole as a set of $N^{1/2}$ heavy
Planckanian constituents with masses $m_P$. In other terms, since
quantum fluctuations inhere in a system at $T\sim 0$, we can regard
a black hole as a set of $N^{1/2}$ Planckanian oscillators in the
ground state
\begin{equation}
\label{mass4}
M=\frac{1}{2G}R_g=\left(\frac{R_g}{l_P}\right)\frac{\hbar
\omega_P}{2}=n\frac{\hbar \omega_P}{2},
\end{equation}
where $n=\left(\frac{R_g}{l_P}\right)\sim N^{1/2}$ and $\omega_P$ is
the Planck frequency, $\omega_P \sim m_P$. Such an interpretation
can have a close relation with the Sakharov's idea of induced
gravity \cite{sa}, \cite{mis}. As Sakharov suggested, the
Einstein-Hilbert action can be induced in a theory with no
gravitational action by integrating out the zero-point fluctuations
of matter fields with Planck frequencies. As is well known, in a
flat spacetime the number of normal modes of vibration per unit
volume in the range of wave numbers from $k$ to $k+dk$ are $\sim
k^{2}dk$. Since each mode of a vacuum oscillation has a zero-point
energy, $\frac{1}{2}\hbar\omega = \frac{1}{2}\hbar c k$, it follows
that the total density of zero-point energy of a matter field
formally diverges as
\begin{equation}
\label{zero}\frac{\hbar}{2}\int^{\infty}_{0}k^{3}dk.
\end{equation}
As Sakharov pointed out, curving spacetime alters it. In a curved
manifold the number of standing waves per unit frequency changes in
such a way that the energy density of vacuum fluctuations becomes
\begin{equation}
\label{den}\rho=A \hbar \int k^{3}dk + B\hbar R\int k dk + \hbar [C
R^{2}+DR^{\alpha\beta}R_{\alpha\beta)}]\int k^{-1}dk
\end{equation}
plus higher-order terms. Here $R$ is the Riemann scalar curvature
invariant, and the numerical coefficients $A,B,...$ are of the order
of magnitude of unity. By means of the renormalization procedure,
the first term in (\ref{den}) must be dropped. The second term,
according to Sakharov, is proportional to the Einstein-Hilbert
action for the gravitational field if effective upper limit in
formally divergent integral in the second term in (\ref{den}) is
taken of the order of magnitude of the reciprocal Planck length,
$k_{cutoff}\sim \omega_P$. The higher order terms lead to
corrections to Einstein's equations and are omitted. Let us now
return to a black hole. Since for a black hole region
$R=(R_{\alpha\beta\gamma\delta}R^{\alpha\beta\gamma\delta})^{1/2}=(12
R_g^{2}/r^{6})^{1/2}$, it follows that the energy of vacuum
fluctuations inside the radius $R_g$ is
\begin{equation}
\label{en}E\sim\rho R_g^{3}=(\hbar R\int_0^{\omega_P} k
dk)R_g^{3}\sim\left(\frac{R_g}{l_P}\right)\frac{\hbar
\omega_P}{2}=n\frac{\hbar \omega_P}{2}=\frac{1}{2G}R_g.
\end{equation}
This coincides with our interpretation of the black hole mass
(\ref{mass4}). In some sense this  answers the question: Why is the
black hole mass linear in the gravitational radius?

Let us now compare thermodynamic fluctuations with quantum ones. As
is easily seen, the thermodynamic fluctuations of the main black
hole quantities (\ref{mass1})-(\ref{ent1}) are considerably larger
than the corresponding quantum fluctuations
(\ref{mass2})-(\ref{ent2}). To be more precise, the mean square
thermodynamic fluctuation equals $N$ times the corresponding quantum
fluctuation,
\begin{equation}
\label{mass3}\langle(\Delta{M})^{2}\rangle_{th}=N
\langle(\Delta{M})^{2}\rangle_{q},
\end{equation}

\begin{equation}
\label{temp3}\langle(\Delta{\beta})^{2}\rangle_{th}=N
\langle(\Delta{\beta})^{2}\rangle_{q},
\end{equation}

\begin{equation}
\label{ent3}\langle(\Delta S)^{2}\rangle_{th}=N \langle(\Delta
S)^{2}\rangle_{q}.
\end{equation}
This resembles a property of another thermodynamic systems (i.e.,
harmonic oscillators) if a mean square quantum fluctuation is taken
as a corresponding thermodynamic fluctuation of one constituent. But
at first sight this looks somewhat strange, because quantum
fluctuations have non-thermal nature. Moreover, as is seen from
(\ref{flu}), quantum fluctuations in geometry of a black hole are
proportional not only to $\hbar$ but also to $T$ (on the other hand,
thermodynamic fluctuations of a black hole (\ref{mass1}) don't
depend on $T$). The fact is that a black hole possesses a very
special property which singles it out; namely, its size and
temperature are not independent parameters. In contrast, quantum
fluctuations of an ordinary harmonic oscillator in its ground state
depend on its energy and don't depend on temperature. Under this
condition fluctuations of a black hole have a dual nature. For
example, we can rewrite the relation (\ref{flu}) as
\begin{equation}
\label{flu2} G^{-1} \Delta R_g \sim T.
\end{equation}
It exhibits the equipartition theorem with $G^{-1}$ playing the role
of the force, $dM/dR_g$. According to the theorem, when a system is
in thermal equilibrium, each of its degrees of freedom contributes
the amount $T/2$ to the total energy. Thus, in addition to
(\ref{mass4}), we have for the black hole mass one more form
\begin{equation}
\label{mass5}M=N\frac{T}{2}~,
\end{equation}
and the specific heat is, therefore, independent of temperature and
equal to $N/2$. But the formula (\ref{mass4}) deals with the number
$n$. Then why don't we use this number instead of $N$, that is, why
not
\begin{equation}
\label{mass6}M=nT?
\end{equation}
Actually we can  do it if we express (\ref{mass6}) in terms of the
local temperature. The point is that the equipartition theorem is
valid only at high temperatures and therefore the formula
(\ref{mass5}) represents the black hole mass at high temperatures;
on the contrary, (\ref{mass4}) expressed in terms $n$ represents the
black hole mass at low temperatures. Thus if we want to use $n$ and
therefore the formula (\ref{mass6}), we should write it as
\begin{equation}
\label{mass7}M=nT_{loc},
\end{equation}
where the local temperature $T_{loc}$ is related with the black hole
temperature (\ref{temp}) by $T_{loc}=T/\chi$, and $\chi$ is the
redshift factor, $\chi=(1-R_g/r)^{1/2}$. We need high temperatures.
Such temperatures are near the horizon. But at $r\rightarrow R_g$
the local temperature diverges. To get rid of the divergence we
should restrict ourselves to some minimum physical distance from the
horizon. If we take $l_P$ as a cutoff so that
$\chi=(1-R_g/r)^{1/2}\approx l_P/2R_g$, we obtain
\begin{equation}
\label{mass7}M=nT_{loc}=n\frac{T}{\chi} \approx
\left(\frac{R_g}{l_P}\right)\frac{T}{\frac{l_P}{2R_g}}\approx
\left(\frac{R_g}{l_P}\right)^{2}T\approx N\frac{T}{2}
\end{equation}
as required. Thus the redshift factor transforms the number of the
effective constituents of a black hole from the value $n$ at
infinity (low temperature limit) to $N$ at the horizon (high
temperature limit).

What is left is to show that the equipartition theorem is valid. As
is well known \cite{land}, it is valid if only the thermal energy
$T$ is considerably larger than the spacing between energy levels.
Is this satisfied in our case? As is well known \cite{land}, the
mean energy spacing of a system is given by
\begin{equation}
\label{main}\delta{E}=\Delta{E}\: e^{-S},
\end{equation}
where $S$ is the entropy of the system and the width $\Delta{E}$ is
some energy interval characteristic of the limitation in our ability
to specify absolutely precisely the energy of a macroscopic system.
It is equal in order of magnitude to the mean fluctuation of energy
of the system. This expression can be immediately applied to a black
hole. Then, taking into account (\ref{ent}) and (\ref{mass1}), we
conclude that
\begin{equation}
\label{fin}T\gg \delta{E}.
\end{equation}
Therefore the equipartition theorem remains valid for the black
holes. Moreover the inequality (\ref{fin}) ensures that quantum
fluctuations of a black hole (\ref{mass2})-(\ref{ent2}) are
negligible. Thus we have removed all doubts about the validity of
the conditions (\ref{cond1}) in the black hole case.

In this paper we have shown that despite a low temperature the
fluctuations of a black hole can be treated thermodynamically.
Moreover, it turns out that the mean square thermodynamic
fluctuation of a black hole equals $N$ times the corresponding
quantum fluctuation, where $N$ can be interpreted as a number of the
effective constituents of a black hole. This can be an evidence of
the complexity of a black hole.


\begin{thebibliography}{99999}

\bibitem{land}
L.~ Landau and E.~ Lifshitz, \emph{Statistical Physics} (Pergamon
Press, Oxford, 1980).

\bibitem{hod}
S.~Hod, Phys. Rev. \textbf{D75}, 064013 (2007).

\bibitem{fro}
V.~Frolov and I.~Novikov, \emph{Black Hole Physics: Basic Concepts
and New Developments} (Kluwer Academic, Dordrecht, 1998).

\bibitem{whe}
J.A.~Wheeler,  \emph{Geometrodynamics} (Academic Press, New York,
1962).

\bibitem{har}
E.~Harrison,  Phys. Rev. \textbf{Dl}, 2726 (1970).

\bibitem{yor}
J.W.~York, Jr., Phys. Rev. \textbf{D28}, 2929 (1983).

\bibitem{ford}
L.H.~Ford and N.F.~Svaiter, Physical Review \textbf{D56}, 2226
(1997).

\bibitem{sa}
A.D.~Sakharov, Sov. Phys. Dokl. \textbf{12},  1040 (1968), reprinted
in Gen. Rel. Grav. \textbf{32}, 365 (2000).

\bibitem{mis}
C.W.~ Misner, K.~Thorn, and J.~Wheeler, \emph{Gravitation} (WH
Freeman and Company, San Francisco, 1973).

\end{thebibliography}
\end{document}